\documentclass{scrartcl}

\usepackage[utf8]{inputenc}
\usepackage[T1]{fontenc}
\usepackage[USenglish]{babel}

\usepackage{mathtools,amsmath,amsthm,amssymb,esint}
\usepackage{float}
\usepackage[unicode=true,pdfusetitle,
 bookmarks=true,bookmarksnumbered=false,bookmarksopen=false,
 breaklinks=true,pdfborder={0 0 1},backref=false,colorlinks=true]
 {hyperref}
\usepackage{siunitx,subfig,cleveref}

\floatstyle{ruled}
\newfloat{algorithm}{tbp}{loa}
\providecommand{\algorithmname}{Algorithm}
\floatname{algorithm}{\protect\algorithmname}

\theoremstyle{plain}
\newtheorem{theorem}{Theorem}
\theoremstyle{remark}
\newtheorem{remark}{Remark}

\DeclareMathOperator{\sgn}{sgn}
\DeclareMathOperator{\linhull}{span}
\DeclareMathOperator{\divergence}{div}

\title{Lagrangian Transport Through Surfaces in Compressible Flows}

\author{Florian Hofherr\thanks{Zentrum Mathematik, Technische Universit\"at M\"unchen, Garching bei M\"unchen, Germany} \and Daniel Karrasch\thanks{Zentrum Mathematik, Technische Universit\"at M\"unchen, Garching bei M\"unchen, Germany (E-mail: \href{mailto:karrasch@ma.tum.de}{karrasch@ma.tum.de})}}

\begin{document}

\maketitle

\begin{abstract}
A material-based, i.e., Lagrangian, methodology for exact integration of
flux by volume-preserving flows through a surface has been developed
recently in {[}Karrasch, \textit{SIAM J.~Appl.~Math.}, 76 (2016), pp.~1178--1190{]}.
In the present paper, we first generalize this framework to general compressible flows, thereby solving the \emph{donating region problem} in full generality.
Second, we demonstrate the efficacy of this approach on a slightly
idealized version of a classic two-dimensional mixing problem: transport in a cross-channel
micromixer, as considered recently in {[}Balasuriya, \textit{SIAM J.~Appl.~Dyn.~Syst.}, 16 (2017), pp.~1015--1044{]}.
\end{abstract}

\section{Introduction}
\label{intro}

Advective transport of conserved quantities is a fundamental physical
process. Its quantification is therefore of importance in a broad
variety of applications. The classic approach proceeds as follows:
first, a surface\footnote{We use the term \emph{surface} as the short form of
\emph{hypersurface}, i.e., a codimension-one submanifold of space or space-time.
In three dimensions, this refers to classic surfaces, in two dimensions to curves.}
of interest is specified, and second, the flux density
is integrated over this surface. The flux density is composed of (i)
the current concentration or density of the scalar quantity, and (ii)
the normal component of the velocity, i.e., the component which is
responsible for transport \emph{across} the surface. This approach
can be viewed as Eulerian, since it looks at transport purely from
the spatial perspective of the surface, ignoring which particles
contribute to the total transport.

This approach comes with drawbacks when complicated geometries need
to be treated \cite{Zhang2013a,Zhang2013b}, or when a restriction
of the flux computation to certain sets of particles, \emph{material}
sets, is of interest. The latter is relevant in the determination
of the physical relevance of \emph{coherent} transport, i.e., transport
by (Lagrangian) coherent structures \cite{Haller2015,Karrasch2016b}, 
see \cite{Dong2014,Zhang2014}
for two transport studies on the global ocean. As was abstractly argued in
\cite{Karrasch2016a}, and as we demonstrate
in \cref{sec:ComparisonFluxEulerianLagrange}, for flux integration
restricted to material sets it is advantageous to
convert the problem of flux integration over surfaces into a Lagrangian
integration problem, because the implementation of the material restriction
is straightforward. The conversion of the flux integration problem has been achieved
rigorously for volume-preserving flows in arbitrary finite dimensions in
\cite{Karrasch2016a} by the second author. 
For such problems, even though theoretically neat and computationally
superior, this approach takes an unusual perspective and may therefore
seem challenging. The aims of this paper are twofold: (i) to generalize the flux integral conversion formula to non-volume-preserving flows, and (ii) to show how to apply the
Lagrangian transport methodology.

This paper is organized as follows. In the rest of this Introduction,
we briefly recall some terminology. In \cref{sec:LagrangeTransport} we
formulate the problem of flux computation and prove our main result, \cref{theorem},
which solves the donating region problem posed by Zhang
\cite{Zhang2013b}; cf.~also Zhang's prior work \cite{Zhang2013,Zhang2015}.
In \cref{sec:2dflows} we specify the general theory to
two-dimensional flow problems and provide an algorithmic overview.
\Cref{sec:ana-example} is devoted to the discussion of an instructive analytic example.
In \cref{sec:micromixer} we discuss extensively a transport problem
in a cross-channel micromixer, that has been inspired by Balasuriya's recent work \cite{Balasuriya2017}.
There, we demonstrate both the conceptual and numerical benefit of a Lagrangian
perspective on transport. We conclude with \cref{sec:conclusion}, and discuss
further implementation details in \cref{sec:Implementation}.

\subsection{Continuum-mechanical terminology}\label{sec:Terminology}

Continuum mechanics considers two frame\-works for the description of
motion and deformation of continua (bodies/fluids) in space, which
are commonly referred to as \emph{Eulerian} and \emph{Lagrangian}.
Eulerian coordinates $x$ are assigned to spatial points in a fixed
frame of reference, Lagrangian coordinates $p$ label material
points and are often taken as the Eulerian coordinates at some initial
time, say, $t=0$. The configuration of the body at time $t$ is given
by the flow map $\varphi_{0}^{t}$, which assigns material points
their location in space, and its motion is given by the one-parameter
family of configurations/flow maps $t\mapsto\varphi_{0}^{t}$.

In fluid dynamics, there are two important characteristic curves associated
with the flow \cite{Batchelor2000}. A \emph{path line through $p$}
is the time-curve of a fixed Lagrangian particle $p$ in Eulerian
coordinates, i.e., $t\mapsto\varphi_{0}^{t}(p)$. In our context, we refer to 
the time-curve of a fixed Eulerian location $x$ in Lagrangian coordinates, i.e.,
$t\mapsto\varphi_{t}^{0}(x)$, as the \emph{streak line through $x$}. In
other words, the streak line is a collection of material points that
will occupy the Eulerian position $x$ at some time. More commonly,
streak lines are viewed as the collection of material points (identified
with the current Eulerian locations) that have passed the Eulerian
position $x$ at some time in the past. It can be imagined as an instantaneous
curve of Lagrangian markers, injected in the past at $x$ and passively
advected by the flow, see \cite{Batchelor2000}.

\section{Lagrangian transport by compressible flows}\label{sec:LagrangeTransport}

Suppose $\varrho$ is conserved by the flow $\varphi$ generated by a time-dependent velocity
field $(x,t)\mapsto\boldsymbol{v}(x,t)=\boldsymbol{v}_t(x)$, i.e., $\varrho$ solves
\begin{align}
\partial_{t}\varrho+\divergence(\varrho\boldsymbol{v}_{t}) & =0, & \varrho(0,\text{\ensuremath{\cdot})} & =\varrho_{0},\label{eq:conservation_law}
\end{align}
where $\varrho_{0}$ is the initial distribution, with no-flux (zero
Neumann) boundary conditions. Then the flux of $\varrho$ across a
stationary surface $\mathcal{C}$ over a time interval $\mathcal{T}=[0,\tau]$
is given by the well-known Eulerian flux integral
\begin{equation}
\int_{\mathcal{T}}\int_{\mathcal{C}}\varrho_t\,\boldsymbol{v}_{t}\cdot\boldsymbol{n}\,\mathrm{d}A\,\mathrm{d}t,\label{eq:Euler_flux}
\end{equation}
where $\boldsymbol{n}$ is the unit normal vector of the surface $\mathcal{C}$
indicating the direction of \emph{positive} flux. This formula generalizes
naturally to smoothly varying surfaces $\hat{\mathcal{C}}=\bigcup_{t\in\mathcal{T}}\lbrace t\rbrace\times\mathcal{C}_{t}$
in space-time
\[
\int_{\hat{\mathcal{C}}}\varrho\,\hat{\boldsymbol{v}}\cdot\hat{\boldsymbol{n}}\,\mathrm{d}\hat{A},
\]
where $\hat{\boldsymbol{v}}=(1,\boldsymbol{v})^{\top}$ is the velocity
field and $\hat{\boldsymbol{n}}$ is the unit normal vector field
of $\hat{\mathcal{C}}$ in space-time (or, extended state space).

In the case of incompressible velocity fields that generate volume-preserving flows,
Karrasch \cite{Karrasch2016a} has shown that the above Eulerian flux integral admits
a Lagrangian equivalent. Our first main result is to show that the assumption of
volume-pre\-ser\-va\-tion by the flow
may be dropped without altering the conclusion. Thus, we solve Zhang's 
\emph{donating region problem} \cite[Def.~1.2]{Zhang2013b} completely.

\begin{theorem}\label{theorem}
For a given regular, time-dependent vector field
$\boldsymbol{v}(x,t)=\boldsymbol{v}_{t}(x)$, let $\varrho(x,t)=\varrho_{t}(x)$ be a
conserved quantity satisfying the Eulerian scalar conservation law \cref{eq:conservation_law}.
Let $\mathcal{C}$ be a compact, connected, embedded codimension-one
surface in (configuration space) $\mathbb{R}^{n}$, and $\mathcal{T}=[0,\tau]$
be a compact time interval. Then there exists a decomposition $\mathcal{D}_{k}\subset\mathbb{R}^{n}$, indexed by $k\in\mathbb{Z}$, of Lagrangian particles, identified by their spatial locations at time $t=0$, that covers a set of full measure, such that
\begin{equation}
\int_{\mathcal{T}}\int_{\mathcal{C}}\varrho_t\,\boldsymbol{u}_t\boldsymbol{\cdot}\boldsymbol{n}
\,\mathrm{d}A\,\mathrm{d}t=(-1)^n \sum_{k\in\mathbb{Z}}\thinspace k\cdot\int_{\mathcal{D}_{k}} \varrho_0(p)\,\mathrm{d}p,\label{eq:aim_of_game}
\end{equation}
where $\boldsymbol{n}$ is the unit normal vector field to $\mathcal{C}$
characterizing the direction of positive flux.
\end{theorem}

\begin{remark}
\Cref{eq:aim_of_game} corrects the corresponding \cite[Eq.~(1)]{Karrasch2016a} by the factor $(-1)^n$.
In \cite{Karrasch2016a} orientation issues have been ignored, such that the given equation there is correct only up to sign; however, it is fully correct for two spatial dimensions, i.e., $n=2$.
\end{remark}

\begin{remark}
As in \cite{Karrasch2016a}, we may generalize \cref{theorem} to smoothly moving surfaces $\hat{\mathcal{C}} = \bigcup_{t\in\mathcal{T}} \mathcal{C}_t$,
i.e., $\hat{\mathcal{C}}$ is everywhere transversal to time fibers $\mathbb{R}^n\times\lbrace t\rbrace$. We omit the straightforward formulation and refer to
\cite[Problem 2]{Karrasch2016a} for the volume-preserving case.
\end{remark}

\begin{remark}
The decomposition $\left(\mathcal{D}_{k}\right)_{k\in\mathbb{Z}}$ is constructed at
the initial time $t=0$. Since it is material, it induces an equivalent decomposition at any other time instance $t\in[0,T]$, in particular at the final time instance.
\end{remark}

We follow the strategy of \cite{Karrasch2016a}, but provide a concise self-contained proof.

\begin{proof}
Let
\begin{align*}
\Psi\colon\mathcal{C}\times\mathcal{T}&\to\mathbb{R}^n, & (x,t)&\mapsto\varphi_t^0(x),
\end{align*}
denote the map assigning to any crossing event $(x,t)$ on the extended surface $\mathcal{C}\times\mathcal{T}$ the corresponding
crossing particle $\varphi_t^0(x)=p$, identified by the initial location of the latter. Here,
$\varphi_a^b$ is the flow map induced by $\boldsymbol{v}$ taking particles from their
spatial location at time $a$ to their spatial location at time $b$. With the flow map at hand, it is 
well-known (see, e.g., \cite[p.~11]{Chorin1993}) that the scalar density $\varrho$ at a later time instance can be 
represented in terms of the initial distribution $\varrho_0$ by
\begin{equation}\label{eq:transport}
\det\left(\mathrm{d}\varphi_0^t\bigr|_{\Psi(x,t)}\right)\varrho_t(x) = \varrho_0(\Psi(x,t)) = \varrho_0(p),
\end{equation}
where $\mathrm{d}f\bigr|_x$ denotes the derivative of $f$ at $x$. In other words, the scalar density is anti-proportional 
(relative to the initial density) to the volume distortion by the flow.
We show in \cref{sec:calculation} that
\begin{equation}\label{eq:calculation}
\boldsymbol{u}_t(x)\boldsymbol{\cdot}\boldsymbol{n}(x) = (-1)^n\det\mathrm{d}\Psi\bigr|_{(x,t)} \cdot\det\mathrm{d}\varphi_0^t\bigr|_{\Psi(x,t)}.
\end{equation}
Combining all together, we have
\begin{align*}
\int_{\mathcal{T}}\int_{\mathcal{C}}\varrho_t\boldsymbol{u}_t\boldsymbol{\cdot}\boldsymbol{n}\,\mathrm{d}A\,\mathrm{d}t &\stackrel{\cref{eq:transport}}{=} 
\int_{\mathcal{T}}\int_{\mathcal{C}} \frac{\varrho_0(\Psi(x,t))}{\det\left(\mathrm{d}\varphi_0^t\bigr|_{\Psi(x,t)}\right)}\boldsymbol{u}_t(x)\boldsymbol{\cdot}\boldsymbol{n}(x)\,\mathrm{d}A(x)\,\mathrm{d}t\\
&\stackrel{\cref{eq:calculation}}{=} (-1)^n\int_{\mathcal{T}}\int_{\mathcal{C}} \varrho_0(\Psi(x,t))\det\mathrm{d}\Psi\bigr|_{(x,t)}\mathrm{d}A(x)\,\mathrm{d}t\\
&= (-1)^n\int_{\mathbb{R}^n} \varrho_0(p)\deg(\Psi,p)\,\mathrm{d}p,
\end{align*}
where
\[
\deg(\Psi,p) = \sum_{(x,t)\in\Psi^{-1}[\lbrace p\rbrace]} \sgn\det\mathrm{d}\Psi\bigr|_{(x,t)}
\]
denotes the (topological) degree of $\Psi$ \cite{Milnor1965} at a regular point $p$, and the 
last equality holds by virtue of a direct consequence \cite[Sect.~3.1.5, Thm.~6]{Giaquinta1998} of the area formula (\cite[Thm.~3.2.3]{Federer1996} and
\cite[Thm.~5.3.7]{Krantz2008}); cf.~also \cite{Karrasch2016a}.

Strictly speaking, the degree function $\deg(\Psi,\cdot)$ may
not be defined on the whole of $\mathbb{R}^n$: excluded are (i) the set of \emph{critical values} of $\Psi$, i.e.,
in our context points $p$ whose trajectory has at least one non-transversal crossing of $\mathcal{C}$ \cite{Karrasch2016a},
and (ii) the image $\Psi[\partial(\mathcal{C}\times\mathcal{T})]$ of the boundary
$\partial(\mathcal{C}\times\mathcal{T})$ of $\mathcal{C}\times\mathcal{T}$ under $\Psi$.
Both sets have measure zero: the first by Sard's theorem, the latter as an 
$(n-1)$-dimensional subset in $\mathbb{R}^n$ \cite[Ch.~6]{Lee2012}.
In other words, the contribution of these particles to transport is negligible.
With the notation as in \cite{Karrasch2016a},
$\mathcal{D}_k \coloneqq \left\lbrace p\in\mathbb{R}^n;~\deg(\Psi,p) = k\right\rbrace$, $k\in\mathbb{Z}$, we may decompose the last integral into
\begin{align*}
\int_{\mathcal{C}\times\mathcal{T}}\varrho_t(x)\boldsymbol{u}_t(x)\boldsymbol{\cdot}\boldsymbol{n}(x)\,\mathrm{d}x\,\mathrm{d}t = \sum_{k\in\mathbb{Z}}\thinspace k\cdot\int_{\mathcal{D}_{k}} \varrho_0(p)\,\mathrm{d}p.
\end{align*}
\end{proof}

\section{Application to two-dimensional flows}\label{sec:2dflows}

\begin{figure}
\centering
\includegraphics{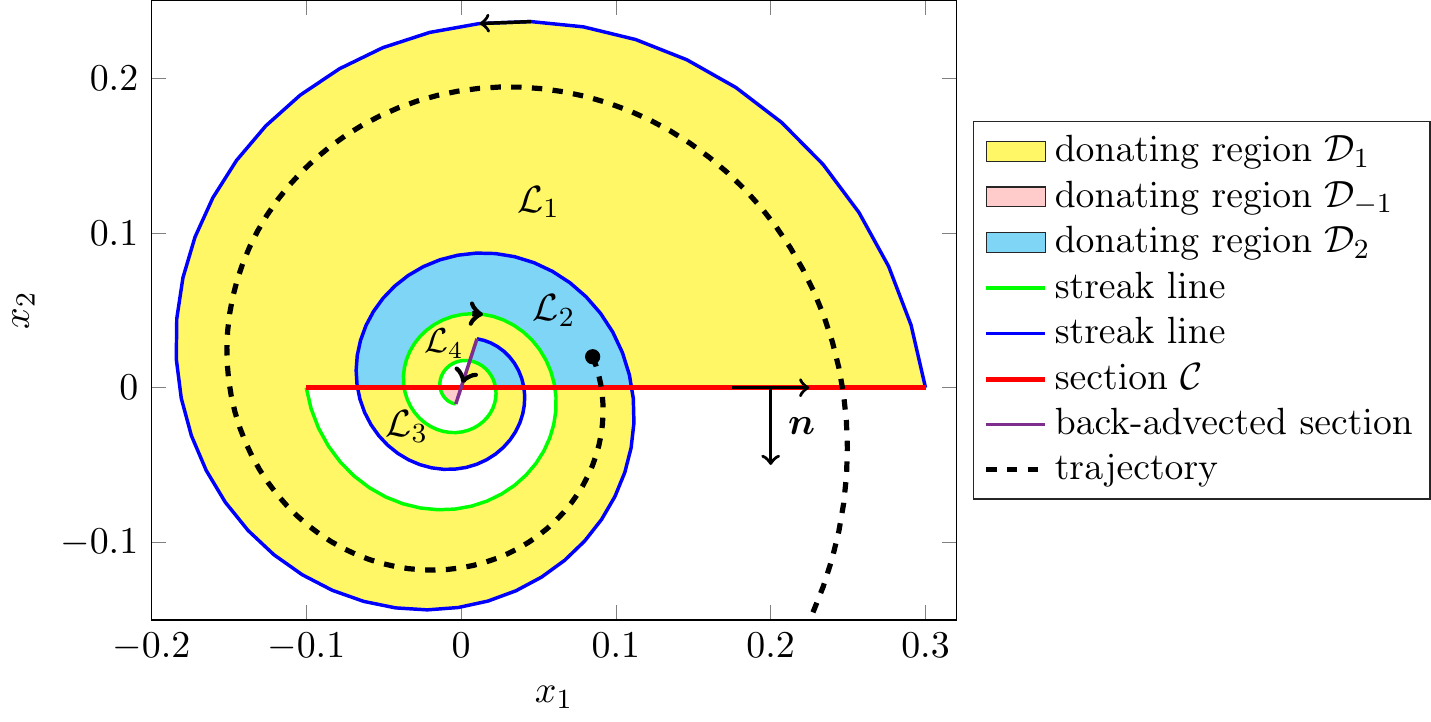}
\caption{Visualization of the donating region $\mathcal{D}$ and its bounding
curve $\partial\mathcal{D}$, corresponding to the example from \cref{sec:ana-example}. The bounding curve is constructed as the concatenation of
section, streak line, back-advected section and the other streak line. Positive
flux is defined to be downward along $\mathcal{C}$ (red). The induced orientation
on $\partial\mathcal{D}$ (indicated by arrows) is such that $\boldsymbol{n}$ points
to the right. The coloring of the loops corresponds to their winding numbers, each 
connected piece corresponds to a simple loop $\mathcal{L}_i$ (not all are labelled
to avoid visual clutter). Enclosed white regions belong to $\mathcal{D}_0$, i.e., 
those particles have a vanishing number of net crossings and do not contribute to
the total transport. The shown trajectory (dashed) originates from $\mathcal{D}_2$ and
has two (positive) crossings of $\mathcal{C}$.}
\label{fig:example}
\end{figure}

As indicated in \cite{Karrasch2016a}, the use of the area formula is to account for
multiple crossings of each particle $p$ from the \emph{donating region} $\mathcal{D}\coloneqq\Psi[\mathcal{C}\times\mathcal{T}]$ as well as the orientation of each crossing.
In two-dimensional flows, bookkeeping of section\footnote{To avoid confusion, we will refer to $\mathcal{C}$ as \emph{section} (and not surface) in the following, since we will be concerned with two-dimensional applications. In this case, $\mathcal{C}$ is a curve.} crossings and their
orientation can be performed easily. It turns out that the net number
of crossings of a given particle $p$, i.e., $\deg(\Psi,p)$, coincides with its \emph{winding
number} $\mathrm{w}_{p}\left(\partial\mathcal{D}\right)$ relative
to the closed curve $\partial\mathcal{D}\coloneqq\Psi[\partial\left(\mathcal{C}\times\mathcal{T}\right)]$ \cite[Sect.~6.6]{Deimling1985}; cf.~\cite{Karrasch2016a,Zhang2013,Zhang2015}.
Generally, $\partial\mathcal{D}$ does not need to be the
topological boundary of the donating region $\mathcal{D}$, but we keep the notation $\partial\mathcal{D}$ by abuse of notation, and refer to it as the \emph{bounding curve} in the sequel. The bounding curve is composed of the section, its back-advected image and the two streak lines emanating from the two endpoints of $\mathcal{C}$; see \cref{fig:example} for a visualization of the involved objects. 
The bounding curve may have a non-trivial topology due 
to self-intersections \cite{Karrasch2016a}.

Recall that the winding number $\mathrm{w}_{p}\left(\partial\mathcal{D}\right)$
of a point $p$ not on $\partial\mathcal{D}$\footnote{There exist different ways
of defining a winding number for points lying on the curve (one of which is
employed in \cite{Zhang2015}), but since the curve has measure/area zero
this does not affect the integral equality, and there is no loss of generality by
omitting this discussion.} relative to $\partial\mathcal{D}$ is
defined as the number of counter-clockwise turns that a vector connecting $p$ to 
$\partial\mathcal{D}$ does, as it follows the bounding curve $\partial\mathcal{D}$
in its positive direction/orientation. The bounding curve is equipped with the
\emph{pullback orientation induced by $\Psi$} \cite[Prop.~15.15]{Lee2012}, i.e., it inherits the orientation of its pre-image $\partial(\mathcal{C}\times\mathcal{T})$; for the latter we
refer the reader to \cref{sec:calculation}. In short, the orientation of $\partial\mathcal{D}$
is such that $\boldsymbol{n}$ points to the right along $\mathcal{C}$ in the plane of particles; cf.~\cref{fig:example}.

As is well-known, the winding number is constant on the interior of simple
loops $\mathcal{L}_{i}$ of the donating region: any two particles that can be connected by
a continuous curve without crossing the bounding curve $\partial\mathcal{D}$ admit the same
winding number, which is why we may introduce the winding number
$\mathrm{w}\left(\partial\mathcal{D},\mathcal{L}_{i}\right)$
of a simple loop $\mathcal{L}_{i}$ relative to $\partial\mathcal{D}$.
Notably, the winding number is fully determined by the bounding curve
alone.

With these considerations, we may re-write \cref{eq:aim_of_game} as
\begin{equation}\label{eq:EulerToLagrange}
\int_{\mathcal{C\times\mathcal{T}}}\varrho(x,t)\boldsymbol{v}_{t}(x)\cdot\boldsymbol{n}(x)\,\mathrm{d}x\,\mathrm{d}t=\sum_{i=1}^{m}\mathrm{w}\left(\partial\mathcal{D},\mathcal{L}_{i}\right)\cdot\int_{\mathcal{L}_{i}}\varrho_0(p)\,\mathrm{d}p.
\end{equation}

As has been mentioned in \cite{Karrasch2016a} and as we demonstrate later, \cref{eq:EulerToLagrange}
is advantageous when transport by a material set $\mathcal{R}$ of interest is to be computed.
Since both the definition of $\mathcal{R}$ and the Lagrangian flux integral
are given in Lagrangian coordinates, the restriction of the integral
to the set $\mathcal{R}$ is direct and explicit. We summarize
the above considerations for two-dimensional flow problems in \cref{alg:Lagrange-approach}.
Some implementation details are given in \cref{sec:Implementation}.

\begin{algorithm}
\caption{Lagrangian transport approach for two-dimensional flow problems}
\label{alg:Lagrange-approach}
\textbf{Input}: Section $\mathcal{C}$, velocity-field $\boldsymbol{v}(x,t)$,
time interval $\mathcal{T}=[0,\tau]$, boundary of the conditioning
set of particles $\partial\mathcal{R}$, distribution of $\rho$ at
$t=0$
\begin{enumerate}
\item Calculate the bounding curve $\partial\mathcal{D}$
\begin{enumerate}
\item Calculate the streak lines through the endpoints of $\mathcal{C}$
\item Calculate the back-advected section
\item Assemble $\partial\mathcal{D}$ by concatenation of (back-advected)
section and streak lines
\end{enumerate}
\item Divide bounding curve $\partial\mathcal{D}$ to obtain simple loops
$\partial\mathcal{L}_{i}$
\begin{enumerate}
\item Find intersections of the bounding polygon
\item Use intersections to assemble simple loops
\end{enumerate}
\item Calculate the winding numbers $\mathrm{w}\left(\partial\mathcal{D},\mathcal{L}_{i}\right)$
\begin{enumerate}
\item Find an interior point of each loop
\item calculate the winding number of this point with respect to the bounding
polygon
\end{enumerate}
\item Evaluate the r.h.s.~of \cref{eq:EulerToLagrange}, potentially restricted to $\mathcal{R}$
\begin{enumerate}
\item Intersect each loop $\mathcal{L}_{i}$ with $\mathcal{R}$
\item Integrate the initial distribution of $\rho$ over the intersections,
or calculate the area of the intersections (if $\rho\equiv1$)
\item Sum up the integrals weighted by the corresponding winding numbers
\end{enumerate}
\end{enumerate}
\textbf{Output}: Accumulated flux due to material originating from $\mathcal{R}$
\end{algorithm}

\section{An analytic example}
\label{sec:ana-example}

We want to test \cref{alg:Lagrange-approach} on a simple example that still shows some nontrivial features
like compressibility and higher winding numbers with mixed signs. To obtain
higher winding numbers, we need particles that cross a given section multiple
times in the \emph{same} direction. It is therefore natural to consider a
divergent rotational linear flow, similar to but simpler than the example in \cite{Karrasch2016a}
and inspired by the tests performed in \cite{Zhang2013a}.

To this end, consider a superposition of rotation
$\Omega=\left(\begin{smallmatrix} 0 & 2\pi\\ -2\pi & 0\end{smallmatrix}\right)$
and strain $S = \left(\begin{smallmatrix} 1 & 0\\ 0 & 1\end{smallmatrix}\right)$, i.e.,
\begin{equation}\label{eq:example}
\dot{\boldsymbol{x}} = \boldsymbol{v}(\boldsymbol{x}) = (\Omega + S)\boldsymbol{x} = \begin{pmatrix} 1 & 2\pi\\ -2\pi & 1\end{pmatrix}\boldsymbol{x},\qquad\boldsymbol{x}=(x_1,x_2),
\end{equation}
with the fluid density
\begin{equation}
\varrho_t(x_1,x_2) = \exp\left(-2t - \exp(-2t)(x_1^2+x_2^2)\right).
\end{equation}
That is we release a Gaussian fluid density $\boldsymbol{x}\mapsto\exp(-(x_1^2+x_2^2))$, centered at the origin, 
at time $0$. It is easily checked that $\varrho$ solves the conservation law \cref{eq:conservation_law}. 

We consider the transport problem for the section $\mathcal{C} = \lbrace -0.1\leq x_1\leq 0.3,~x_2=0\rbrace$
over the time interval $\mathcal{T} = [0,2.2]$, where positive flux is downward. The geometry
of the donating region as computed by
\cref{alg:Lagrange-approach} is shown in \cref{fig:example}. Close to the origin, we 
see a region of winding number -1, which corresponds to particles which cross the
section three times upward (i.e., negative crossings) and twice downward, with a
net number of crossings equal to -1. The enclosed white regions correspond to particles
that cross as often upward as downward, and therefore their transport contribution
annihilates. The region with winding number 2 corresponds to particles that cross
$\mathcal{C}$ twice more often in the positive than in the negative direction. This is
achieved by reflux from below that is bypassing the section $\mathcal{C}$; see the shown trajectory (dashed).

\begin{table}
\centering
\caption{Analysis of accuracy and computation time (measured on one core of an Intel Core i5 2.3 GHz processor) for the Lagrangian 
transport computation, depending on a distance threshold used in the adaptive construction of streak lines and the back-advected section.}
\label{tab:accuracyLagrange}
\begin{tabular}{|c|c|c|}
\hline
Distance threshold & Relative error & Computation time [s]\\\hline
0.05 & 3.18e-03 & 0.39\\\hline
0.01 & 1.44e-04 & 0.56\\\hline
0.005 & 4.87e-06 & 0.92\\\hline
0.001 & 2.75e-07 & 4.16\\\hline
\end{tabular}
\end{table}

The relative error to the analytic result and the computation times for our implementation of
\cref{alg:Lagrange-approach} are given in \cref{tab:accuracyLagrange},
depending on a threshold for the maximal distance of consecutive points along
the streak lines and the back-advected section. This threshold is used in their adaptive
calculation: the smaller the threshold, the finer resolved are streak lines and
back-advected section; see \cref{sec:Implementation} for more details. A
thorough numerical analysis of a similar algorithm has been performed by Zhang in
\cite{Zhang2013a}.

\section{Transport in a cross-channel micromixer}\label{sec:micromixer}

\subsection{Set-up}
\label{sec:set-up}

\begin{figure}
\centering
\includegraphics{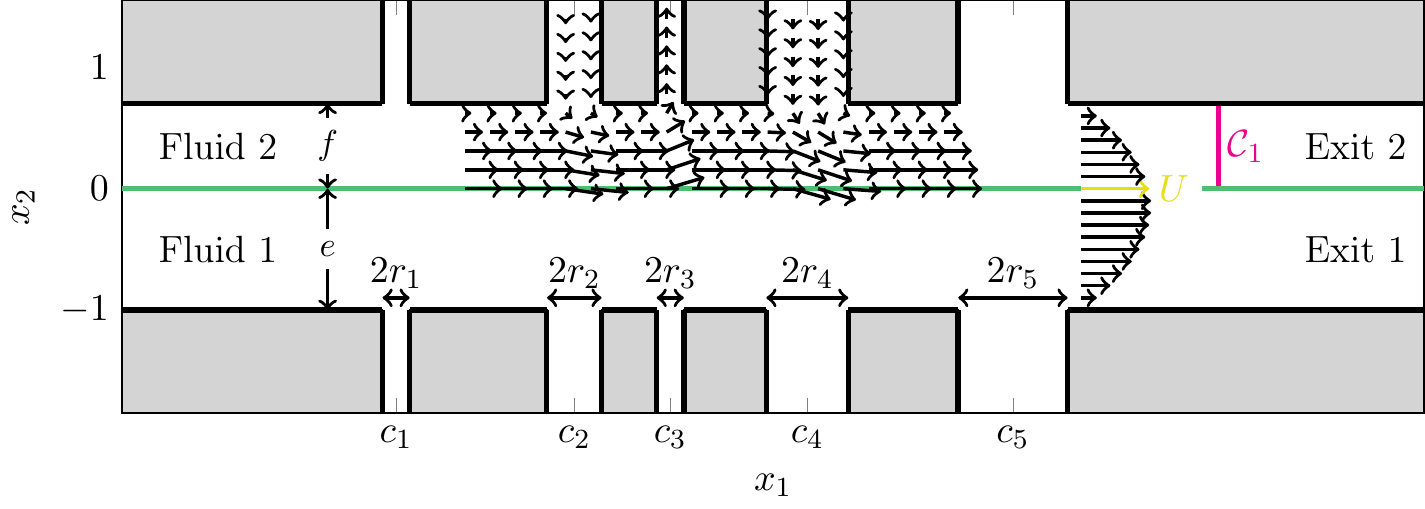}
\caption{Setup of a cross-channel micromixer with 5 cross channels.
The two fluids enter from the left and flow to the right. At the end of the channel,
the unperturbed background velocity profile is
shown. In the upper half, the full instantaneous velocity field 
at some instance in time is indicated. For further details, refer to the text.}
\label{fig:Setup}
\end{figure}

We consider the following two-di\-men\-sio\-nal setup shown in \cref{fig:Setup},
which is strongly inspired by \cite{Balasuriya2017}; cf.~also \cite{Tabeling2004}.
Two incompressible fluids enter a straight channel on the left side, fluid
1 in the lower region and fluid 2 in the upper region of the channel.
In the absence of an external disturbance the two fluids remain separated
by their common interface indicated by the horizontal line $x_2=0$ (green). Physically,
the only possible mixing is due to diffusion along the interface,
which is, however, not as efficient as often desired. To improve mixing for
microfluidic applications, cross-channels are introduced, that are
disturbing the horizontal flow in the vertical direction by periodic
sucking and pumping. The resulting apparatus is called a \emph{cross-channel
micromixer} \cite{Tabeling2004}. Consistently with \cite{Balasuriya2017}, we neglect
diffusive transport in the following, and discuss the validity of this assumption in \cref{sec:conclusion}.

We assume that the velocity field is modeled as the superposition of a steady
horizontal background field and a time-dependent, $\varepsilon$-modulated vertical
cross-channel velocity agitation, i.e.,
\[
\boldsymbol{v}_{t}(x_1,x_2)=\boldsymbol{v}(t,x_1,x_2)=\boldsymbol{v}_{\textnormal{c}}(x_2)+\varepsilon\sum_{i=1}^{5}\boldsymbol{v}{}_{i}(t,x_1).
\]
For the flow in the main channel, we assume a parabolic flow-pattern
with flow speed $U$ at the interface and no-slip boundary conditions
at the walls. The corresponding velocity profile is shown on the right
of \cref{fig:Setup}. For the given geometric parameters, the steady
background velocity field takes the form
\[
\boldsymbol{v}_{\textnormal{c}}(x_2)=U\left(-\frac{1}{ef}x_2^{2}+\left(\frac{1}{e}-\frac{1}{f}\right)x_2+1\right)\boldsymbol{e}_{1},
\]
where $e$ and $f$ are the widths of fluid 1 and fluid 2, respectively, see \cref{fig:Setup}.
The assumed profile for the cross-channel flows is likewise parabolic
and reads as
\begin{equation}
\boldsymbol{v}_i(t,x_1)=
\begin{cases}
\frac{v_i}{r_i^2} \left((x_1-c_i)^2-r_i^2\right)\cos(\omega t+\phi_i)\boldsymbol{e}_2, & \lvert x_1-c_i\rvert\leq r_i,\\
0, & \textnormal{else}.
\end{cases}
\end{equation}
Here, $v_{i}$ is the maximal velocity magnitude at the center of
the $i$-th channel, $r_{i}$ is half of the width of the channel,
$c_{i}$ is the position of the center of the channel, $\omega$ is
the frequency of the vertical disturbance and $\phi_{i}$ represents
the phase shift. The flow in the main channel is supposed to be dominant,
i.e., we have $|\boldsymbol{v}{}_{i}|\leq U$ and $0\leq\varepsilon<1$.
An exemplary part of the superposed velocity field is shown in the
upper half of the mixer in \cref{fig:Setup} for some instant of time.

Due to the (assumed) absence of diffusive mixing, we will quantify \emph{mixing}
of the two fluids under the unsteady velocity agitation by the amount
of fluid 1 that leaves cross-channel micromixer via the upper part
of the channel (exit 2) which would be occupied by fluid 2 without
the velocity agitation (and vice versa). Consequently, quantifying
mixing is equivalent to quantifying \emph{transport}, i.e.,\emph{
accumulated flux}, of fluid 1 across the section $\mathcal{C}_{1}$,
see \cref{fig:Setup}.

\subsection{The Eulerian transport approach}

\subsubsection{Choice of section and other issues}\label{sec:Choice-of-section}

In light of Eq.~\eqref{eq:Euler_flux}, it may seem that the question of
quantifying transport of fluid 1 through exit 2 can be answered directly
by applying the flux integral formula to some section like $\mathcal{C}=\mathcal{C}_{1}$,
that connects a point on the original fluid interface behind the cross-channel region
with the upper channel wall; see \cref{fig:Setup}. This way, however, we would measure
the out-flux of both fluid 1 and fluid 2 through exit 2, which is
a useless assessment of mixing in the micromixer.

There are two possible attempts to fix this: (i) choose a different
section $\mathcal{C}$ to distinguish the two fluids according to
their origin, and/or (ii) multiply the fluid density $\varrho(\boldsymbol{x},t)$
by a characteristic function $\chi_{\textnormal{fluid }1}(\boldsymbol{x},t)$
corresponding to the region occupied by fluid 1 in space over time.

Regarding the first option, a natural candidate would be the initial
interface itself, bounded by a point $a$ left of the cross-channel
section and a point $b$ right of it, with normal direction $\boldsymbol{e}_{2}$;
that is, the portion of the black dashed line between $a$ and $b$
in \cref{fig:StreaklineApproach}. The rationale behind such a choice
is that any material carried by fluid 1 which leaves the micromixer
through the upper exit 2 has to cross this section. Generally, this is
a minimal requirement for a meaningful choice of section for this problem:
essentially any section connecting $b$ to $a$
or $b$ to the upper wall has to be crossed by fluid 1 on the way to exit 2
and is therefore admissible.
The issue here is, however, that with a direct application of the Eulerian flux integral
to admissible sections we would again measure any flux of both
fluid 1 and fluid 2. Also, it does not help to restrict attention
to the positive flux only, i.e., replacing $\boldsymbol{v}_{t}(\boldsymbol{x})\cdot\boldsymbol{n}(\boldsymbol{x})$
by $\max\{\boldsymbol{v}_{t}(\boldsymbol{x})\cdot\boldsymbol{n}(\boldsymbol{x}),0\}$, because
this way we would neglect reflux of fluid 1 but would take into account
reflux of fluid 2 that has invaded the lower region before.

In contrast, the second option would in principle do the job with
any choice of section just discussed, but is numerically extremely
challenging. While it is simple to decide at time $t=0$ which spatial
region is occupied by fluid 1 and which one by fluid 2, this knowledge
is expensive to obtain for later times as it corresponds to solving
numerically the conservation law \eqref{eq:conservation_law}. Alternatively,
one may proceed reversely: choose an admissible section $\mathcal{C}$ and discretize
$\mathcal{C}\times\mathcal{T}$ to apply some numerical quadrature
scheme to solve \cref{eq:Euler_flux}. For each point $(\boldsymbol{x},t)\in\mathcal{C}\times\mathcal{T}$,
compute its past flow image at the initial time to see whether the
particle $p$ occupying $\boldsymbol{x}$ at time $t$ originated from fluid 1
or fluid 2. Correspondingly, its instantaneous flux value is, respectively,
included and excluded from the numerical quadrature scheme. This kind
of numerical quadrature turns out to be expensive even for short time
intervals $\mathcal{T}$, see \cref{sec:ComparisonFluxEulerianLagrange}.

\subsubsection{The streak line approach\label{sec:streak_line_approach}}

\begin{figure}
\centering
\includegraphics{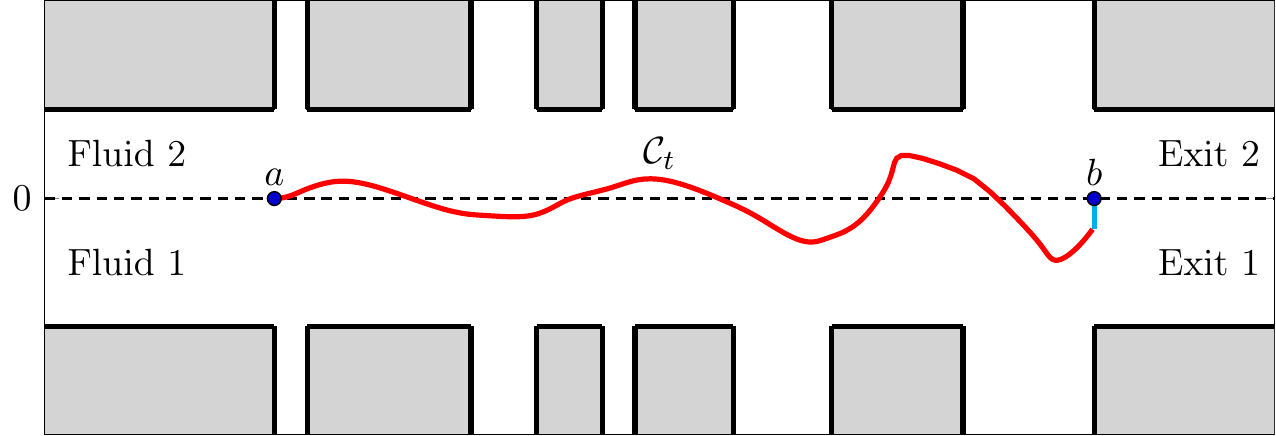}
\caption{Construction of the nominal fluid interface $\mathcal{C}_{t}$ at
a given time $t$, following \cite{Balasuriya2017}: the
upstream streak line through $a$ (red) and the vertical gate (cyan) at $b$.
The original interface for the undisturbed case is indicated by the horizontal thin
dashed line. Particles located above the downstream streak line belong to fluid 2.
The shown situation corresponds to an instantaneous flux of fluid 2 into the exit 1 region.}
\label{fig:StreaklineApproach}
\end{figure}

Recently, Balasuriya \cite{Balasuriya2017} proposed a more involved
construction of a time-de\-pen\-dent section $\mathcal{C}_{t}$
for a flow problem that is very related to the one described in \cref{sec:set-up}.
He considers two anchor points $a$ and $b$ on the original fluid interface, which
are located before and after the cross-channel region, respectively;
see \cref{fig:StreaklineApproach}.

At any time instance $t$, the so-called upstream streak line of $a$
(red) is considered, i.e., a collection of Eulerian positions of material
points which have passed through $a$ before time $t$. This corresponds to the classic notion of streak lines eluded to in \cref{sec:Terminology}. 
The streak line is continued up to the point when it hits a line perpendicular to the original fluid interface and attached to $b$. The \emph{nominal fluid interface} $\mathcal{C}_{t}$ is then constructed by a concatenation of upstream streak
line and a straight, vertical \emph{gate} (blue) connecting the hitting point with $b$; see \cite{Balasuriya2017}.
Finally, one builds a continuously varying section in space-time by
taking the union of all nominal fluid interfaces $\mathcal{C}_{t}$.
This section can be considered as half-Eulerian and half-Lagrangian,
since it is build partially from a material curve, the streak line, and
partially from a spatially defined curve, the gate.

By construction, the streak line segment of the nominal fluid interface
behaves like a material curve and, therefore, admits no flux.
Therefore, the upstream streak line forms the fluid interface throughout,
and depending on whether it hits the $x_1=b$ line above or below $b$, some of fluid
1 is leaving through exit 2 or fluid 2 through exit 1. In the first case, the
gate has an orientation such that flux to the right is measured positively,
in the second case flux is measured negatively. By restricting to one sign
of flux only, one may compute transport, say, of fluid 1 through
exit 2.

Since the streak line segment of the nominal interface $\mathcal{C}_{t}$ is material,
any flux across $\mathcal{C}_{t}$ is across the gate. The aim of \cite{Balasuriya2017}
is to expand the (instantaneous) flux $\phi$ across the gate, i.e.,
\[
\phi(t)=\int_{\mathcal{C}_{t}}\varrho_t\,\boldsymbol{v}_{t}\cdot\boldsymbol{n}\,\mathrm{d}A
\]
in powers of the perturbation strength parameter $\varepsilon$ (implicit
in $\boldsymbol{v}_{t}$) and to determine the leading order term\footnote{By construction, the nominal interface $\mathcal{C}_{t}$ has only
tangential variation over time, which is why the (normal component
of the) relative velocity equals the original velocity $\boldsymbol{v}_{t}$.}.
This is a feasible approach provided that the streak line has a unique
intersection with the $x_1=b$ line, which
can be satisfied by choosing $\varepsilon$ sufficiently small. In
cases when the streak line has multiple intersections with a horizontal
line (say, an S-shaped pattern), the gate would consist of several
pieces, and the perturbative approach breaks down.

\subsection{The Lagrangian transport approach}\label{sec:Lagrange}

In this section, we demonstrate the efficacy of the Lagrangian transport formalism on 
the cross-channel micro\-mi\-xer with 5 cross-channels
described in \cref{sec:set-up}. The parameters are given in \cref{tab:ParametersExperiments}.
For the experiments, we assume that a steady flow\textemdash with
the cross-channels turned off\textemdash is established in the main
channel before the experiment starts. Thus, fluid 1 occupies the lower,
fluid 2 the upper region, and both are separated by their mutual interface
at $x_2=0$. Possible turbulence due to the sharp edges at the cross-channels
is neglected. At time $t=0$ the cross-channels are switched on and
the mixing process begins.  For convenience, the initial
distribution of the conserved quantity is set to $\varrho_0\equiv1$
throughout this section.

\begin{table}[tbhp]
\caption{Parameters for the cross-channel mixer. The strength of the disturbance
$\varepsilon$ is chosen separately for each experiment.}
\label{tab:ParametersExperiments}
\centering
\begin{tabular}{|lcc|}
\hline 
Parameter & Variable & Value(s)\tabularnewline
\hline 
Velocity at the interface of the channel  & $U$  & 1\tabularnewline
\hline 
Height occupied by the lower fluid  & $e$  & 1\tabularnewline
\hline 
Height occupied by the upper fluid  & $f$  & 0.7\tabularnewline
\hline 
Positions of the centers of the cross-channels  & $c_{i}$  & $[1,\,2.3,\,3,\,4,\,5.5]$ \tabularnewline
\hline 
Vertex-velocities of the cross-channels  & $v_{i}$  & $[1,\,0.5,\,0.3,\,0.8,\,1]$\tabularnewline
\hline 
Half-widths of cross-channels  & $r_{i}$  & \emph{$[0.1,\,0.2,\,0.1,\,0.3,\,0.4]$}\tabularnewline
\hline 
Phase shift between cross-channels  & $\phi_{i}$  & $\pi\cdot[1,\,2,\,3,\,4,\,3.5]$\tabularnewline
\hline 
Sucking/pumping frequency  & $\omega$  & 4\tabularnewline
\hline 
\end{tabular}
\end{table}

\subsubsection{Visual evaluation of donating region}\label{sec:Graphical-Evaluation}

First, we want to evaluate graphically how well the bounding curve
$\partial\mathcal{D}$ captures the particles that are crossing the
chosen section. To this end, we choose the section $\mathcal{C}$
as the straight line along the initial interface from $a$ at the
start of the first cross channel to $b$ at the end of the last cross
channel. We set $\varepsilon=1$, which is considered as huge in the
perturbative setting of \cite{Balasuriya2017}.

\begin{figure}
\centering
\subfloat[]{\includegraphics{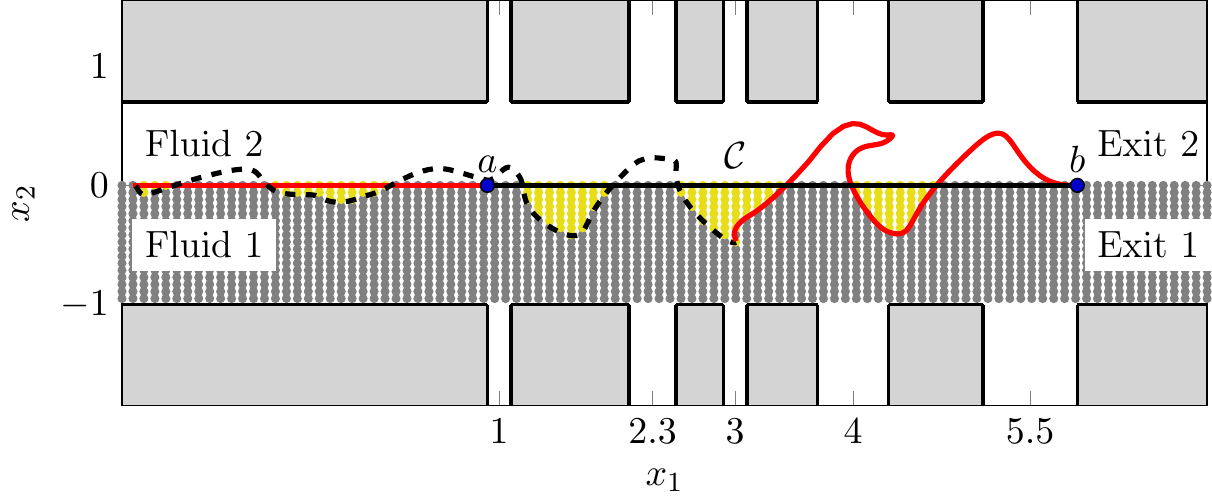}
\label{fig:deltaD_and_particles}}

\subfloat[]{\includegraphics{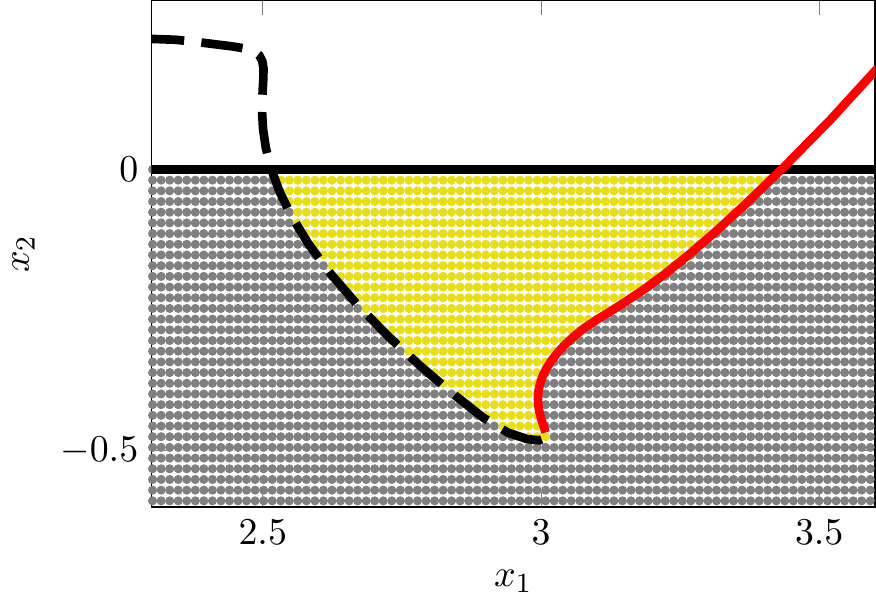}
\label{fig:deltaD_and_particles_enlaged}}
\caption{(a) Particles in their initial positions colored depending on whether
they end up above (yellow) or below (gray) the initial interface after
the time interval $[0,3]$. The bounding curve $\partial\mathcal{D}$
is composed of section (black), streak lines (red) and back-advected
section (dashed). (b) Close-up of \cref{fig:deltaD_and_particles}.}
\label{fig:deltaD}
\end{figure}

Specifically, we compute the path lines for a grid of fluid 1 type
particles over the time interval $\mathcal{T}=[0,3]$. If the end
position of a particle has a positive $y$-component, the particle
has crossed $\mathcal{C}$ and, therefore, contributes to the transport
within the time interval. If it has a negative $y$-component, it does
not contribute to transport. In \cref{fig:deltaD}, the particles are
plotted in their initial positions and colored depending on whether
they end up above (yellow) or below (gray) the initial interface.
Independently, the first step of \cref{alg:Lagrange-approach} is performed
with 200 discretization points for $\mathcal{C}$ and $\mathcal{T}$
each. The resulting bounding curve of the donating regions is plotted
in the same figure as the particles.

The results of the two independent experiments match perfectly, see
\cref{fig:deltaD_and_particles}. All crossing sample particles (yellow)
are enclosed correctly by $\partial\mathcal{D}$. Note that the loops
of the bounding curve that are located above the initial interface
contain all the fluid-2 type particles that end up in the lower half.
By increasing the particle resolution we again find perfect matching,
see \cref{fig:deltaD_and_particles_enlaged}.

An animation of this simulation experiment is provided in 
the Supplementary Material SM1.

\subsubsection{Comparison of Eulerian and Lagrangian transport approaches}
\label{sec:ComparisonFluxEulerianLagrange}

Next, we compare computational results and effort for evaluating the
Lagrangian and the Eulerian versions of the flux integral. To this
end, we set $\varepsilon=1$ and aim at computing the transport of
fluid 1 across the horizontal section described in \cref{sec:Graphical-Evaluation}
over the time interval $\mathcal{T}=[0,2]$.

For the Eulerian flux integral an indicator function is used for the
distinction of the two fluids, as detailed in \cref{sec:Choice-of-section}.
Since $\boldsymbol{n}\equiv(0,1)^{T}$, the term $\boldsymbol{v}\cdot\boldsymbol{n}$
simply yields the $\boldsymbol{e}_{x_2}$-component of $\boldsymbol{v}$.
The characteristic function $\chi_{\text{fluid }1}$ is evaluated by
back-advection of points $(\boldsymbol{x},t)$ of some grid discretization
of $\mathcal{C}\times\mathcal{T}$ back to the initial time $t=0$,
and subsequently checking if the $y$-component of the result is less
than 0. In this case, $\chi_{\text{fluid }1}(\boldsymbol{x},t)$ equals
1, otherwise it is 0. Finally, a numerical quadrature based on the
trapezoidal rule was applied to compute the integral non-adaptively,
for increasing numbers of integration points.

For the Lagrangian approach, $\mathcal{C}$ and $\mathcal{T}$ were
discretized initially by 100 equidistant points each, and
then refined adaptively, to yield 168 and 201 points, respectively;
see \cref{sec:Implementation} for details on the adaptive integration
scheme. The conditioned set of particles $\mathcal{R}$ restricting
the transport calculation to fluid 1 is specified by a sufficiently
large rectangular polygon $\partial\mathcal{R}$ enclosing the lower
half of the channel.

\begin{table}
\centering
\caption{Analysis of accuracy and numerical effort for the conditional Eulerian transport computation, depending on the number of integration points.}
\label{tab:ComparisonFlux}
\begin{tabular}{|c|c|c|}
\hline
No.~of integration points & Relative error & Integration time [s]\\\hline
1k & 1.4744e-01 & 0.5\\\hline
5k & 7.1727e-02 & 0.8\\\hline
10k & 4.9097e-02 & 1.2\\\hline
25k & 2.5313e-02 & 2.2\\\hline
50k & 1.7816e-02 & 3.4\\\hline
75k & 1.4843e-02 & 4.5\\\hline
100k & 1.1801e-02 & 5.5\\\hline
250k & 7.3630e-03 & 10.9\\\hline
500k & 4.9478e-03 & 20.6\\\hline
750k & 3.7236e-03 & 30.4\\\hline
1,000k & 3.3845e-03 & 37.9\\\hline
2,000k & 2.1410e-03 & 76.9\\\hline
3,000k & 1.7187e-03 & 102.4\\\hline
5,000k & 1.2459e-03 & 171.1\\\hline
10,000k & 7.3614e-04 & 364.5\\\hline
30,000k & 2.6525e-04 & 1036.8\\\hline
\end{tabular}
\end{table}

The results are shown in \cref{tab:ComparisonFlux}. The numerical quadrature
of the Eulerian integral converges to the result of the Lagrangian
calculation, the relative error decreases from approx.~15\% (i.e.,
one correct digit) for 10\textsuperscript{3}=1k integration points to
0.026\% (4 correct digits) for 3$\cdot$10\textsuperscript{7} integration
points. Thus, many sampling points are required to produce reasonably
accurate results. The computation times\footnote{The timings are measured on an Intel Core i5 (dual core) with 2.3
GHz.} increase from about 1 second for 10\textsuperscript{3} up to more than
1,000 seconds for 3$\cdot$10\textsuperscript{7} integration points.
For comparison, the Lagrangian computation takes less than 2 seconds. In
fact, in our specific example in the perturbed situation, transport
can occur only in the range of the cross-channels. Therefore, it would
be sufficient to apply numerical quadrature only to those parts of
the section. This would roughly halve the computational effort; of
course, such information is not available, in general.

\subsubsection{Different choices of sections}

\begin{figure}
\centering
\includegraphics{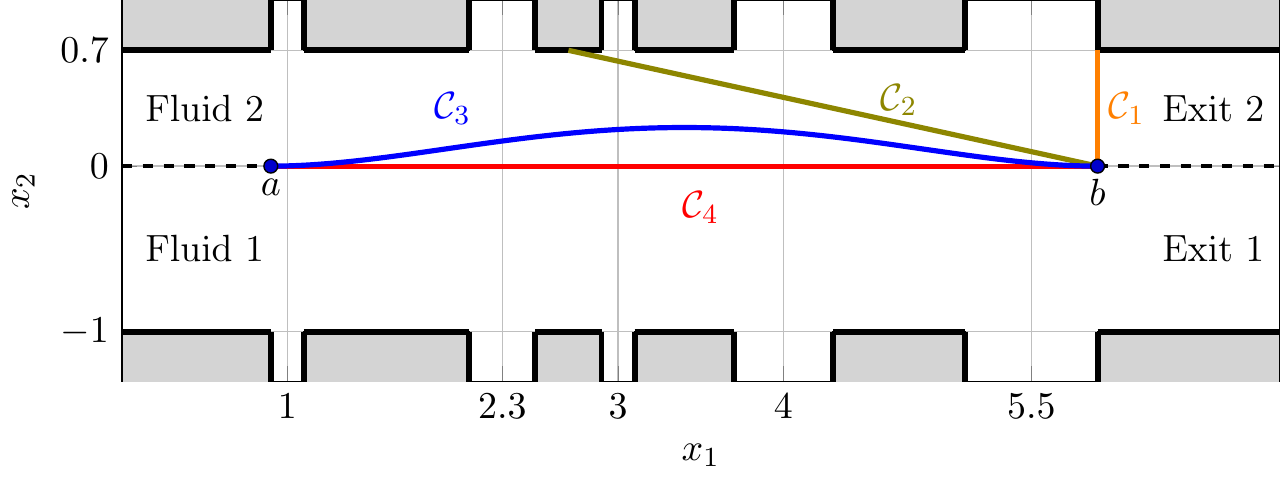}
\caption{Different choices of sections. The point $a$ is at the $x_1$-position
of the beginning of the first cross-channel, $b$ is at the $x_1$-position
of the ending of the last cross-channel.}
\label{fig:Setup_DifferentSections}
\end{figure}

Finally, we compare the transport calculation for different choices
of sections to demonstrate that any section constructed according
to the criteria discussed in \cref{sec:Choice-of-section} measure essentially
the same transport; see \cref{fig:Setup_DifferentSections} for the considered
sections. The point $a=(0.9,0)$ is at the beginning of the first
channel and $b=(5.9,0)$ is at the end of the last channel and therefore
all sections are admissible.
For all sections, the accumulated flux of fluid 1 is determined for
different time intervals $\mathcal{T}=[0,\tau_{i}]$ using the Lagrangian
approach. The restriction to fluid 1 is achieved as in \cref{sec:ComparisonFluxEulerianLagrange}.

We choose several values for $\tau_{i}$ from $\tau=1$ to $\tau=6.2$
in $0.2$-steps. This implies that the time intervals $\mathcal{T}$
become large which---for large $\varepsilon$---leads
to possibly poor approximations of the donating region. To ensure
proper computation, we set $\varepsilon=0.3$.

\begin{figure}
\centering
\subfloat[]{\includegraphics{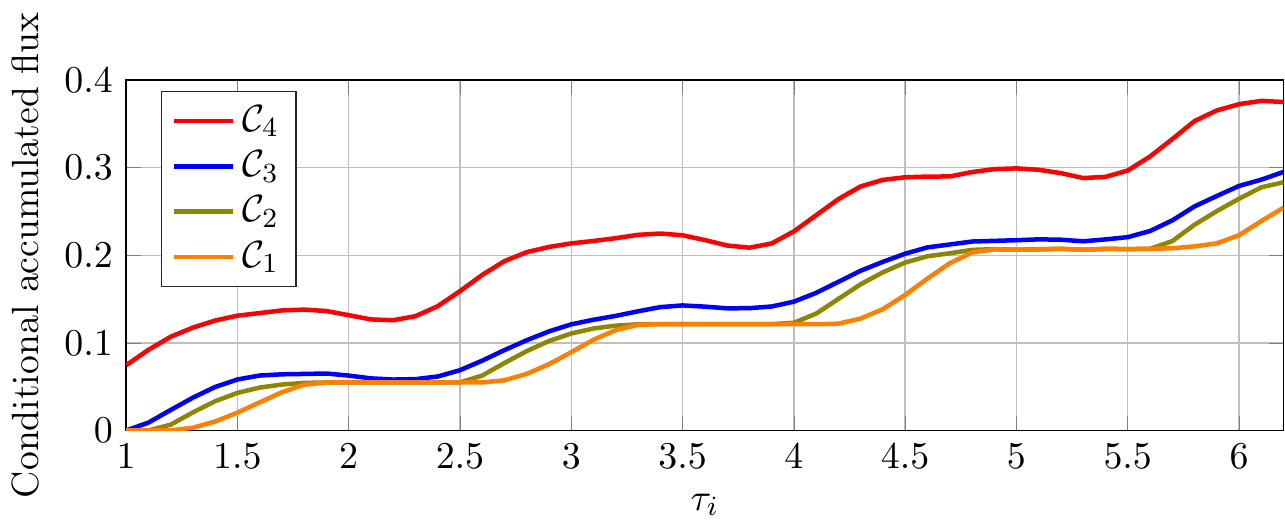}
\label{fig:transport_over_time}}\\

\subfloat[]{\includegraphics{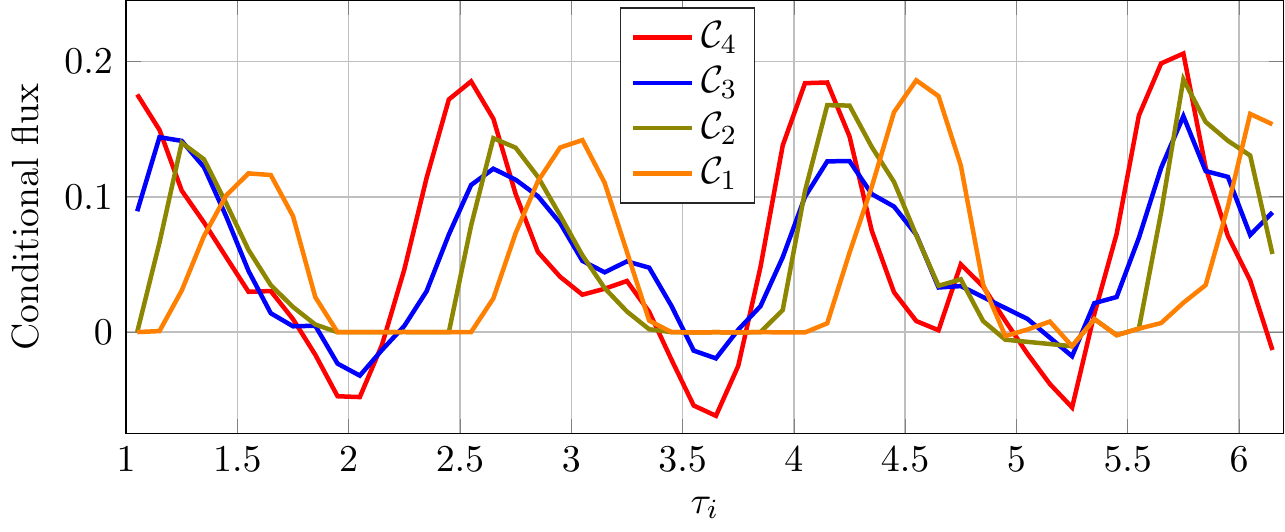}
\label{fig:flux_over_time}}
\caption{Transport of fluid 1 across the sections $\mathcal{C}_{k}$ from \cref{fig:Setup_DifferentSections}.
(a) Accumulated flux over the time intervals $\mathcal{T}=[0,\tau_{i}]$.
(b) Instantaneous flux at $\tau_{i}$, obtained from finite-differencing
of the accumulated flux function in (a).}
\label{fig:time_evolution_fluxes}
\end{figure}

The resulting conditional accumulated fluxes are shown in \cref{fig:transport_over_time}.
As expected, the lines follow the same trend and an offset for the
different sections is noticeable, which is, however, not constant
over time. For each $\tau_{i}$, the transport across section $\mathcal{C}_{4}$
is maximal and the transport across $\mathcal{C}_{1}$ is minimal.
This is due to the fact that fluid 1 has to cross $\mathcal{C}_{4}$
first anyway before getting to $\mathcal{C}_{1}$. Therefore, the
difference between the two curves at $\tau_{i}$ is exactly the amount
of fluid that has crossed $\mathcal{C}_{4}$ but not yet $\mathcal{C}_{1}$
by the time $\tau_{i}$. The other offsets are explained analogously.

To get a finer intuition, consider the evolution of conditional fluxes
across the sections, shown in \cref{fig:flux_over_time}. As before,
we are conditioning on flux of fluid 1 only. These curves are obtained
from finite-differencing the conditional accumulated flux curves shown in
\cref{fig:transport_over_time},
and are therefore a short-term flux approximation to the actual instantaneous
conditional flux densities. As expected, section $\mathcal{C}_{4}$
is showing the strongest reflux, i.e., negative flux, of fluid 1 downward,
since it is everywhere normal to the velocity agitation in the cross-channels.
Up to small numerical inaccuracies around $\tau_{i}\approx5.2$, when
the boundaries of the donating region become extremely complicated,
the flux across $\mathcal{C}_{1}$ is always non-negative. This is
trivial when taking an Eulerian viewpoint, since the velocity field
is---at any time instance and at any point on $\mathcal{C}_{1}$---pointing
rightwards (or vanishing at the wall), irrespective of whether fluid
1 or fluid 2 is crossing. From the Lagrangian viewpoint, recall that
our computations are based on the area of induced donating regions.

We provide animations of the evolution of the donating regions over increasing time
intervals for $\mathcal{C}_{1}$ (SM2) and $\mathcal{C}_{4}$ (SM3) for the
parameters used in this section in the Supplementary Material.

\begin{figure}
\centering
\includegraphics{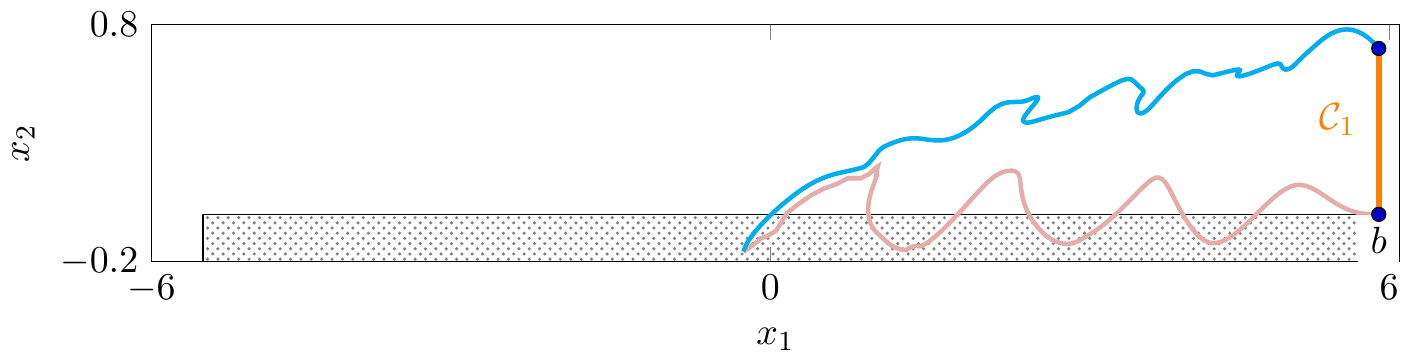}\\[1.5em]

\includegraphics{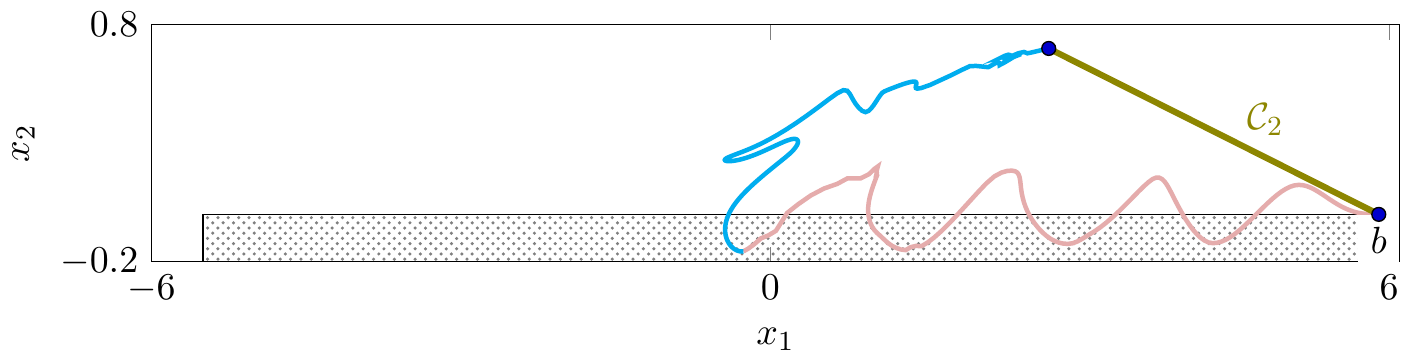}\\[1.5em]

\includegraphics{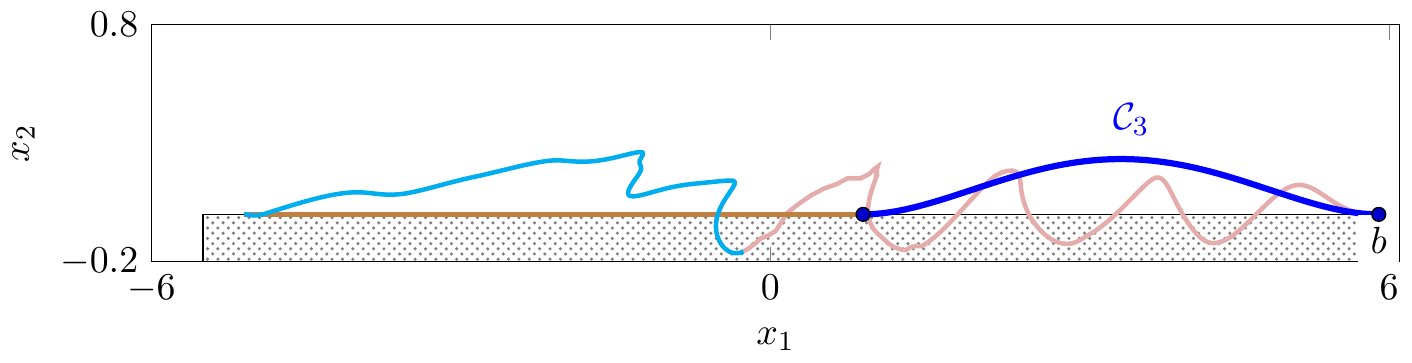}\\[1.5em]

\includegraphics{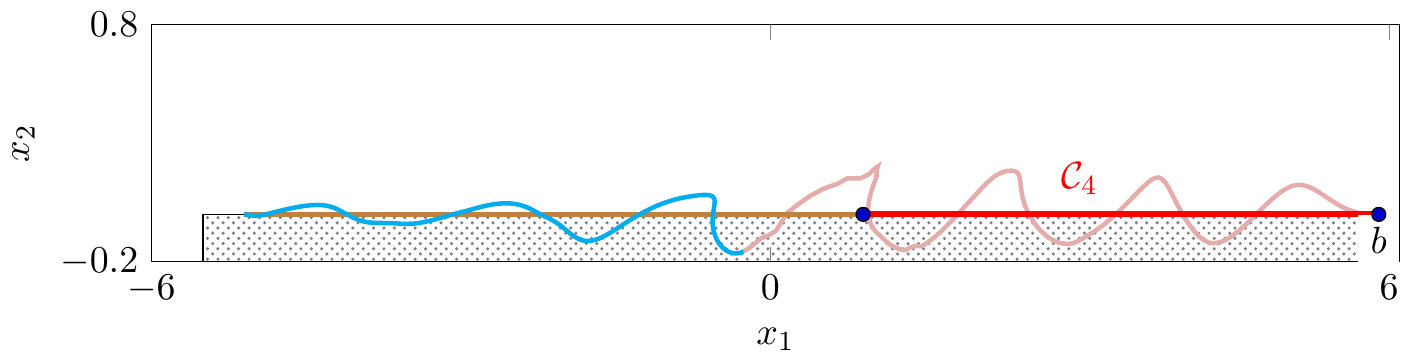}
\caption{Components of $\partial\mathcal{D}$ for the sections shown in 
\cref{fig:Setup_DifferentSections} and for integration time $\tau=6$: the sections (colored as in \cref{fig:Setup_DifferentSections}),
the two streak lines (pink and brown) and the back-advected section (cyan).
The region of origin is shown by
the dotted area. Note that for $\mathcal{C}_{1}$ and $\mathcal{C}_{2}$
one of the streak lines degenerates to a point, for $v=0$ in the respective
upper endpoint of those sections.}
\label{fig:CompareBoundingPolygonsDifferentSections}
\end{figure}

To see that the different sections indeed capture essentially the
same transport, consider \cref{fig:CompareBoundingPolygonsDifferentSections}.
For all sections, the constituents of $\partial\mathcal{D}$ are shown
for $\tau=6$: The back-advected section (cyan), the two streak
lines (brown and pink) and the section itself (in the color corresponding
to \cref{fig:Setup_DifferentSections}). The restricting region of origin
is shown by the dotted rectangular area. In the Lagrangian approach, the transport
of fluid 1 to exit 2 is calculated by integrating the initial distribution
of $\varrho_0$ over the simple loops of $\partial\mathcal{D}$ that
have a non-empty intersection with the dotted area.

Now consider point $b$. As it is the endpoint of all four sections,
its corresponding streak line (pink) is exactly the same. Thus, this
part of the bounding curve yields the same contribution to the transport
for all cases. The difference in the accumulated flux arises from
the varying back-advected surfaces (cyan). The contribution to the
transport by this part of the bounding polygon is the highest for
$\mathcal{C}_{4}$ and the lowest for $\mathcal{C}_{1}$ which matches
the results shown in \cref{fig:time_evolution_fluxes}. So we can quantify
the differences between the curves by calculating the contribution
of the cyan part of the bounding polygon to the transport. As the
shape of the back-advected section is not constant for different $\tau_{i}$,
this contribution is neither, which explains the non-constant differences
between the conditional transport curves.

The above discussion applies verbatim to the time-dependent
nominal interface $\mathcal{C}_{t}$ introduced by Balasuriya in \cite{Balasuriya2017} and
described in \cref{sec:streak_line_approach}.
Here, at any time instance both endpoints are stationary and coincide
with $a$ and $b$, i.e., the ones of $\mathcal{C}_{3}$ and $\mathcal{C}_{4}$.
For any flow interval, the streak line parts of the boundary of the
donating regions generated by $\mathcal{C}_{3}$, $\mathcal{C}_{4}$
or $\mathcal{C}_{t}$ are the same, and the only (time-dependent)
differences are due to the final sections and their back-advected images (for
$\mathcal{C}_{3}$ and $\mathcal{C}_{4}$) on the one hand, and the initial
$\mathcal{C}_{0}$ and final $\mathcal{C}_{\tau}$ nominal interfaces on the other
hand. Up to these explainable differences due to the exact choice of section, 
the same transport is measured by the Lagrangian transport approach and the
Eulerian flux across the (possibly split) gate from the streak line approach \cite{Balasuriya2017}.

\section{Concluding remarks}\label{sec:conclusion}

We have extended the Lagrangian transport approach from the volume-preserving
flow case \cite{Karrasch2016a} to the general case. Also, we demonstrated both the 
efficacy and accuracy of the Lagrangian approach to transport across surfaces in a 
cross-channel micromixer problem. As argued theoretically in \cite{Karrasch2016a}, this approach
is advantageous when the interest in the computation of transport
(integrated flux) is restricted to a specific material set. Roughly
speaking, instead of determining which points on the surface are occupied
by material points of interest and subsequently computing the conditional Eulerian
transport integral as usual, we transform the Eulerian transport
integral into a Lagrangian one, whose restriction to the material
set of interest is straightforward.

Consistently with \cite{Balasuriya2017}, we have neglected diffusive transport
in the micromixer problem. This assumption, however, appears to be rather
unrealistic. As discussed in detail in \cite{Nguyen2004} (and the references therein),
diffusion \emph{is} present, and may be even dominating advection as measured via the
P\'{e}clet number. Nevertheless, it is often \emph{not
efficient enough} for the desired mixing purpose. This is why mixing by diffusion is
further enhanced by the use of complicated channel geometries or cross-channel velocity
agitations, or, generally speaking, by \emph{chaotic advection},
\cite{Aref2002}. The inclusion of the effect of diffusion turns the conservation law
\cref{eq:conservation_law} into an advection-diffusion equation, which is no longer
of conservation/transport type. Still, a (generalized) conservation law can be formulated,
to which our framework can be applied again, and work is in progress in this direction.

For another instructive comparison of the classic Eulerian with the
Lagrangian perspective, consider the animations from the Supplementary
Material, SM1 and SM3. In both cases, the exact same process as an
evolution in time is shown: from the Eulerian
perspective (SM1), in which particles are moving through space, colored
according to whether they end up in the upper half at the end of the
observation time interval; and from the Lagrangian perspective (SM3),
where particles are colored according to whether they have crossed the horizontal section
by the current time. Effectively, the fact
that they have crossed the section is represented through coloring,
not by drawing them in their current spatial position.

The Matlab implementation of \cref{alg:Lagrange-approach}, which is available on
github\footnote{https://github.com/dkarrasch/FluxDoRe2D}, is robust. In our many
numerical experiments, only extremely thin filaments and near-tangent
line segments appeared to be challenging for the winding number computation,
which are generally known challenges in (time-dependent) polygon computations.
Future work on the computational front will be concerned with implementations
for three-dimensional flows.

\appendix

\section{Proof of \Cref{eq:calculation}}\label{sec:calculation}

We prove
\[
\boldsymbol{u}_t(x)\boldsymbol{\cdot}\boldsymbol{n}(x) = (-1)^n\det\mathrm{d}\Psi\bigr|_{(x,t)} \cdot\det\mathrm{d}\varphi_0^t\bigr|_{\Psi(x,t)}.
\]
To this end, we need to first specify orientations; for an introduction to orientations, see, e.g., \cite[Ch.~15]{Lee2012}.

For definiteness, we equip space $\mathbb{R}^n$ and time $\mathbb{R}$ with their respective standard orientations \cite[Example 15.2]{Lee2012}. Next, we equip $\mathbb{R}^n\times\mathbb{R}$ with the product orientation, cf.~\cite[Prop.~15.7]{Lee2012}: let $(\boldsymbol{e}_1,\ldots,\boldsymbol{e}_n)$ be spatially (positively) oriented, then $(\boldsymbol{e}_1,\ldots,\boldsymbol{e}_n,\boldsymbol{e}_t)$ is oriented in space-time. For example, in two spatial dimensions time points upward as in the space-time figures in \cite{Karrasch2016a}.

We equip the extended surface $\mathcal{C}\times\mathcal{T}$ with an orientation induced by the unit normal vector field $\boldsymbol{n}$ as follows \cite[Prop.~15.21]{Lee2012}.
An ordered basis $(\boldsymbol{e}_2,\ldots,\boldsymbol{e}_n,\boldsymbol{e}_t)$ of the tangent space of $\mathcal{C}\times\mathcal{T}$ is oriented if and only if $(\boldsymbol{n},\boldsymbol{e}_2,\ldots,\boldsymbol{e}_n,\boldsymbol{e}_t)$ is oriented in space-time. Finally, we equip the boundary of $\mathcal{C}\times\mathcal{T}$ with the \emph{Stokes orientation} \cite[p.~386]{Lee2012} by means of an outward-pointing vector field $\boldsymbol{v}$: an ordered basis $(\boldsymbol{e}_2,\ldots,\boldsymbol{e}_n)$ of the boundary tangent space is oriented if and only if $(\boldsymbol{v},\boldsymbol{e}_2,\ldots,\boldsymbol{e}_n)$ is oriented in $\mathcal{C}\times\mathcal{T}$, which holds if and only if $(\boldsymbol{n},\boldsymbol{v},\boldsymbol{e}_2,\ldots,\boldsymbol{e}_n)$ is oriented in space-time.

For instance, when $n=2$, at the ``lower'' boundary $\mathcal{C}\times\lbrace 0\rbrace$, the negative time vector $-\boldsymbol{e}_t$ is an outward-pointing vector, and the orientation along $\mathcal{C}\times\lbrace 0\rbrace\cong\mathcal{C}$ is given by $\boldsymbol{e}$ such that $(\boldsymbol{n},-\boldsymbol{e}_t,\boldsymbol{e})$ is positively oriented in three-dimensional space-time. A simple right-hand check reveals that then $\boldsymbol{n}$ points to the right of $\boldsymbol{e}$, or, equivalently, that $(\boldsymbol{n},\boldsymbol{e})$ is a positively oriented basis in space $\mathbb{R}^2$. Indeed, if $(\boldsymbol{n},-\boldsymbol{e}_t,\boldsymbol{e})$ is oriented in space-time, then so is $(\boldsymbol{n},\boldsymbol{e},\boldsymbol{e}_t)$, and by the above definition of the space-time orientation, it follows that $(\boldsymbol{n},\boldsymbol{e})$ is oriented in space.\footnote{Since $\Psi$ acts on $\mathcal{C}\times\lbrace 0\rbrace$ like the identity map, the induced orientation
on $\partial\mathcal{D}$ is the one for which $\boldsymbol{n}$ points to the right along $\mathcal{C}$.}

Next, fix a regular point $(x,t)\in\mathcal{C}\times\mathcal{T}$. Then $\Psi$
acts diffeomorphically between an open neighborhood $\mathcal{U}$
of $(x,t)$ in $\mathcal{C}\times\mathcal{T}$ and its image $\mathcal{V}\coloneqq\Psi(\mathcal{U})$ in $\mathcal{D}\subseteq\mathbb{R}^n$. We may choose local coordinates on $\mathcal{U}$
such that in $(x,t)$ we have a positively oriented orthonormal basis, where $\boldsymbol{e}_{1}=\boldsymbol{n}\in T_{x}^{\perp}\mathcal{H}$, $\linhull\left\{\boldsymbol{e}_{2},\ldots,\boldsymbol{e}_{n}\right\} =T_{x}\mathcal{C}$, and $\boldsymbol{e}_{t}\in T_{t}\mathcal{T}$ the (unit) positive time direction. The following parallelepipeds
\begin{gather}
\bigwedge_{i=1}^n\boldsymbol{e}_{i}=\boldsymbol{e}_{1}\wedge\ldots\wedge\boldsymbol{e}_{n} \qquad \text{in space, and}\label{eq:space}\\
\boldsymbol{e}_{2}\wedge\ldots\wedge\boldsymbol{e}_{n}\wedge\boldsymbol{e}_{t} \qquad\text{on the extended surface,}\label{eq:surface}
\end{gather}
span (positive) unit volume parallelepipeds, whose volume we also denote by the wedge product for notational simplicity.

On the one hand, we have
\begin{equation}\label{eq:lhs}
\boldsymbol{u}_{t}\boldsymbol{\cdot}\boldsymbol{n}=\boldsymbol{u}_{t}\boldsymbol{\cdot}\boldsymbol{e}_1=\det\begin{pmatrix} | & | & & |\\ \boldsymbol{u}_{t} & \boldsymbol{e}_{2} & \cdots & \boldsymbol{e}_{n}\\ | & | & & |\end{pmatrix} = \boldsymbol{u}_{t} \wedge \boldsymbol{e}_{2} \wedge \cdots \wedge \boldsymbol{e}_{n}
= \boldsymbol{u}_{t} \wedge\bigwedge_{i=2}^n\boldsymbol{e}_{i},
\end{equation}
where the second equality holds true including the sign because of \cref{eq:space}.
The determinant is an $(n\times n)$-determinant, where all vectors can be
represented, for example, in the $(\boldsymbol{e}_{1},\ldots,\boldsymbol{e}_{n})$-basis.
On the other hand, we first observe that
\[
\det\mathrm{d}\varphi_{0}^{t}\bigr|_p\det\mathrm{d}\Psi\bigr|_{(t,x)}=\det\left(\mathrm{d}\varphi_{0}^{t}\bigr|_p\mathrm{d}\Psi\bigr|_{(t,x)}\right),
\]
due to the multiplicativity of the determinant. The last determinant is again an $(n\times n)$-determinant, where the image vectors of $(\boldsymbol{e}_2,\ldots,\boldsymbol{e}_n,\boldsymbol{e}_t)$ can be represented in the $(\boldsymbol{e}_{1},\ldots,\boldsymbol{e}_{n})$-basis.
Thus, we may compute $\det\mathrm{d}\varphi_{0}^{t}\bigr|_p\det\mathrm{d}\Psi\bigr|_{(t,x)}$
as the change of volume under the composed action of $\mathrm{d}\varphi_{0}^{t}\bigr|_p\mathrm{d}\Psi\bigr|_{(t,x)}$ on the oriented normalized parallelepiped $\boldsymbol{e}_{2}\wedge\ldots\wedge\boldsymbol{e}_{n}\wedge\boldsymbol{e}_{t}$ (recall \cref{eq:surface}):
\begin{equation}\label{eq:rhs}
\bigwedge_{i=2}^n\left(\mathrm{d}\varphi_{0}^{t}\bigr|_p\mathrm{d}\Psi\bigr|_{(t,x)}\boldsymbol{e}_{i}\right)\wedge\left(\mathrm{d}\varphi_{0}^{t}\bigr|_p\mathrm{d}\Psi\bigr|_{(t,x)}\boldsymbol{e}_{t}\right)
=\bigwedge_{i=2}^n\boldsymbol{e}_{i}\wedge\mathrm{d}\varphi_{0}^{t}\bigr|_p\boldsymbol{w}_{t}
= \bigwedge_{i=2}^n \boldsymbol{e}_{i}\wedge - \boldsymbol{u}_{t},
\end{equation}
where we have used (i) that $\mathrm{d}\varphi_{0}^{t}\bigr|_p\mathrm{d}\Psi\bigr|_{(t,x)}$ acts like the identity map on tangent vectors $\boldsymbol{e}_i$, $i=2,\ldots,n$, of $\mathcal{C}$, and (ii) an identity for the \emph{streak vector field}
\[
\boldsymbol{w}_t\coloneqq\mathrm{d}\Psi\boldsymbol{e}_t=\partial_t\phi_t^0(x)=-\left(\mathrm{d}\phi_0^t\right)^{-1}\boldsymbol{u}_t,
\]
which was proven in \cite[Sec.~3]{Karrasch2016a}. To finish the proof, it remains to switch the sign of $-\boldsymbol{u}_t$ and to shift it to the first position in \cref{eq:rhs}, which requires $1+(n-1)=n$ reversions in orientation, hence the $(-1)^n$-coefficient.

\section{Implementation details\label{sec:Implementation}}

\paragraph{Step 1: Construction of $\partial\mathcal{D}$}

We use Matlab's \texttt{ode45} adaptive fourth-order Run\-ge-Kut\-ta ODE
solver to compute streak lines directly from their definition and
for the back-advected section. Additionally, we implemented an adaptive
point insertion scheme to ensure that (i) points along the streak
lines and the back-advected section are not too far apart and that
(ii) consecutive line segments do not form an angle too far away from
the straight one. At the same time, point insertion is prevented when
consecutive points are already sufficiently close. This is to prevent
self-overlapping zigzag-patterns in regions of high numerical sensitivity.

\paragraph{Step 2: Detection of intersections and of simple loops}

Currently, self-inter\-sec\-tions of the bounding polygon are computed
by checking pairwise line segments for crossings. The intersection
points are inserted into the polygon, and simple loops are extracted
by going ``as left as possible'' at intersections in a counter-clockwise
orientation, until one arrives at the starting point. A further decrease
in computational effort could be achieved through the sweepline methodology
\cite{Berg2008}.

\paragraph{Step 3: Construction of interior points and computation of winding
numbers}

Once simple loops have been computed, the next aim is to determine
each individual winding number $\mathrm{w}\left(\partial\mathcal{D},\mathcal{L}_{i}\right)$.
To this end, we need to construct a representative interior point
$p$ and compute its winding number classically by counting the number
of turns of a line segment connecting $p$ to consecutive points along
$\partial\mathcal{D}$. For robustness, we construct many candidates
for interior points, slightly on the left of each line segment of
$\mathcal{L}_{i}$ in a counter-clockwise parametrization. If all
those winding numbers coincide, we take it as $\mathrm{w}\left(\partial\mathcal{D},\mathcal{L}_{i}\right)$.
Problems with this construction may occur in thin filaments, where
the little left-offset may already yield an exterior point. In such a
case, we construct the skeleton of the simple polygon and use its
branching points as interior point candidates. As before, if their
computed winding numbers all coincide we take it as $\mathrm{w}\left(\partial\mathcal{D},\mathcal{L}_{i}\right)$.

\paragraph{Step 4: Integration of the Lagrangian flux formula}

For uniform initial material densities and volume-preserving flows, the
integration of $\varrho(0,\cdot)$ over each simple loop $\mathcal{L}_{i}$
reduces essentially to the computation of its respective area, for which Matlab
admits the fast built-in function \texttt{polyarea}.

For general non-uniform initial material densities a numerical quadrature of
$\varrho_0$ has to be performed on each $\mathcal{L}_{i}$. There are at least two
fundamental approaches to this. First, one may triangulate/mesh $\mathcal{L}_{i}$
and then apply Gauss quadrature of some chosen degree on each triangle.
There exist Matlab functions such as \texttt{quadpts} from the iFEM package \cite{Chen2008},
which give the location of the quadrature points in a triangle (in barycentric coordinates)
and their corresponding weights. Second, there are theory and Matlab
packages available that perform Gauss quadrature directly on polygons, PolyGauss \cite{Sommariva2007}.
For convex sets, the quadrature points are guaranteed to lie inside
the domain of integration; for non-convex sets this may no longer
be true, though, see \cite{Sommariva2007}.

\section*{Acknowledgements}

We would like to thank Christian Ludwig for useful discussions and for support for his
Matlab suite VISUALCOMPLEX. D.K.~also acknowledges stimulating discussions with J\"org Schumacher.


\end{document}